\documentclass[fleqn,twoside]{article}
\usepackage{espcrc2}

\usepackage[centertags]{amsmath}
\allowdisplaybreaks[1]
\usepackage{amsbsy}
\usepackage{amsfonts}
\usepackage{amssymb}

\usepackage{graphicx}
\usepackage{dcolumn}
\usepackage{bm}

\usepackage{cite}
\usepackage{url}

\title{Short-BaseLine Electron Neutrino Disappearance}

\author{
Carlo Giunti
\address{INFN, Sezione di Torino, and Department of Theoretical Physics, University of Torino,
Via P. Giuria 1, I--10125 Torino, Italy}
and
Marco Laveder
\address{Dipartimento di Fisica ``G. Galilei'', Universit\`a di Padova,
and
INFN, Sezione di Padova,
Via F. Marzolo 8, I--35131 Padova, Italy}
}

\begin{document}

\begin{abstract}
We analyzed
the electron neutrino data of the Gallium radioactive source experiments
and
the electron antineutrino data of the reactor Bugey and Chooz experiments
in terms of neutrino oscillations.
We found a hint of
a CPT-violating asymmetry of the effective neutrino and antineutrino mixing angles.
\end{abstract}

\maketitle

The
GALLEX
and
SAGE
radioactive source experiments
revealed a disappearance of electron neutrinos
with energy $E$ of the order of 1 MeV at a distance $L$ of the order of 1 m
which could be due to short-baseline oscillations
\cite{Laveder:2007zz-brief,hep-ph/0610352,0707.4593,0711.4222,0902.1992,1005.4599,1006.2103,1006.3244,1008.4750}.

\begin{figure}[t!]
\includegraphics*[bb=22 144 572 704, width=\linewidth]{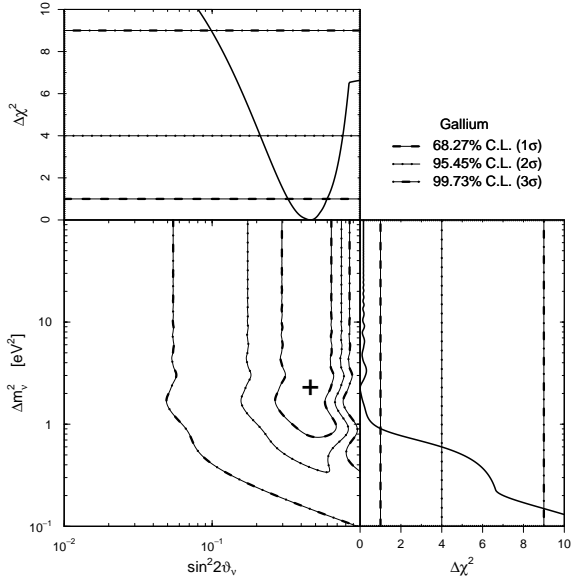}
\caption{ \label{027}
Results of the fit of the data of the Gallium radioactive source experiments
\cite{1006.3244}.
The best-fit point is indicated by a cross.
}
\end{figure}

We considered the effective short-baseline (SBL) electron neutrino survival probability
\begin{equation}
P_{\nu_{e}\to\nu_{e}}^{\text{SBL}}(L,E)
=
1
-
\sin^2 2\vartheta_{\nu}
\sin^2\!\left( \frac{ \Delta{m}^2_{\nu} L }{ 4 E } \right)
\,,
\label{026}
\end{equation}
where
$\vartheta_{\nu}$ is the effective neutrino mixing angle and
$\Delta{m}^2_{\nu}$ is the effective neutrino squared-mass difference.
We found
the best-fit values \cite{1006.3244}
\begin{equation}
\sin^2 2\vartheta_{\nu,\text{bf}} = 0.46
\,,
\quad
\Delta{m}^2_{\nu,\text{bf}} = 2.24 \, \text{eV}^2
\,.
\label{028}
\end{equation}
Figure~\ref{027} shows the allowed regions
in the
$\sin^{2}2\vartheta_{\nu}$--$\Delta{m}^{2}_{\nu}$ plane
and
the
marginal $\Delta\chi^{2} = \chi^2 - \chi^2_{\text{min}}$'s,
from which one can infer the uncorrelated allowed intervals
of
$\sin^{2}2\vartheta_{\nu}$ and $\Delta{m}^{2}_{\nu}$.

\begin{figure}[t!]
\includegraphics*[bb=22 144 572 704, width=\linewidth]{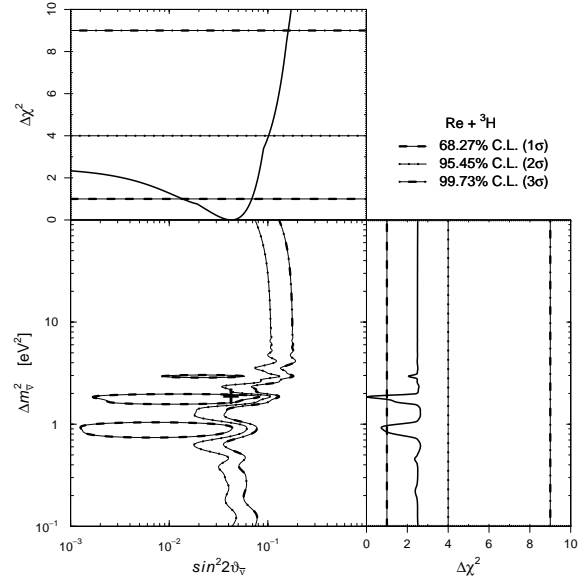}
\caption{ \label{040}
Results of the fit of
the data of reactor and Tritium $\beta$-decay experiments
\cite{1005.4599}.
The best-fit point is indicated by a cross.
}
\end{figure}

Considering antineutrinos,
a fit of the data of the Bugey and Chooz
reactor antineutrino experiments
in terms of the
effective short-baseline electron antineutrino survival probability
\begin{equation}
P_{\bar\nu_{e}\to\bar\nu_{e}}^{\text{SBL}}(L,E)
=
1
-
\sin^2 2\vartheta_{\bar\nu}
\sin^2\!\left( \frac{ \Delta{m}^2_{\bar\nu} L }{ 4 E } \right)
\,,
\label{026a}
\end{equation}
where
$\vartheta_{\bar\nu}$ is the effective antineutrino mixing angle and
$\Delta{m}^2_{\bar\nu}$ is the effective antineutrino squared-mass difference,
gives the
best-fit values \cite{1005.4599}
\begin{equation}
\sin^2 2\vartheta_{\bar\nu,\text{bf}} = 0.042
\,,
\quad
\Delta{m}^2_{\bar\nu,\text{bf}} = 1.85 \, \text{eV}^2
\,.
\label{028a}
\end{equation}
Figure~\ref{040} shows the allowed regions
in the
$\sin^{2}2\vartheta_{\bar\nu}$--$\Delta{m}^{2}_{\bar\nu}$ plane
and
the
marginal $\Delta\chi^{2}$'s,
obtained taking into account also the constraints on the mixing given by the results of
the
Mainz
and
Troitsk
Tritium $\beta$-decay experiments \cite{1005.4599}.

CPT symmetry implies that the survival probabilities of neutrinos and antineutrinos are equal
(see Ref.~\cite{Giunti-Kim-2007}),
i.e.
$\sin^{2}2\vartheta_{\nu}=\sin^{2}2\vartheta_{\bar\nu}$
and
$\Delta{m}^{2}_{\nu}=\Delta{m}^{2}_{\bar\nu}$.
Figs.~\ref{027} and \ref{040}
show that $\sin^{2}2\vartheta_{\nu}$ is likely to be larger than about 0.1,
whereas $\sin^{2}2\vartheta_{\bar\nu}$ is likely to be smaller than about 0.1.
The incompatibility of neutrino and antineutrino data
in the case of CPT symmetry is quantified by a
0.2\%
parameter goodness-of-fit \cite{1008.4750}.
Hence, we have a hint
of CPT violation in short-baseline $\nu_{e}$ and $\bar\nu_{e}$ disappearance
which could be complementary to that found recently
in the MINOS long-baseline $\nu_{\mu}$ and $\bar\nu_{\mu}$
disappearance experiment \cite{MINOS-Neutrino2010}.

\begin{figure}[t!]
\includegraphics*[bb=22 147 564 698, width=\linewidth]{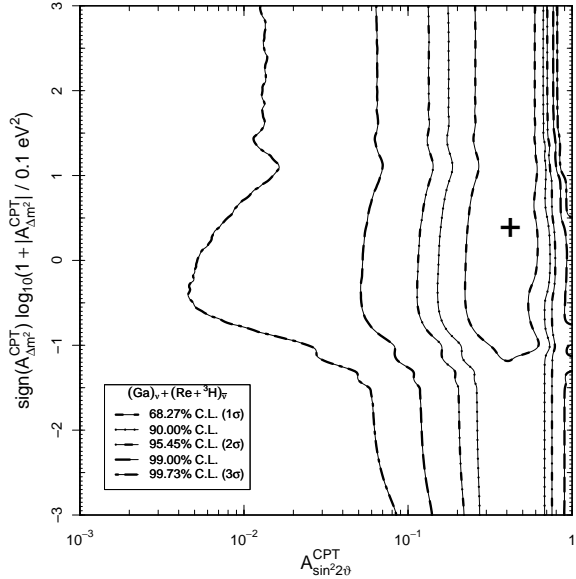}
\caption{ \label{006b}
Allowed regions in the
$A_{\sin^22\vartheta}^{\text{CPT}}$--$A_{\Delta{m}^{2}}^{\text{CPT}}$ plane
\cite{1008.4750}.
The best-fit point is shown by a cross.
}
\end{figure}

Analyzing
the Gallium data
and
the reactor plus Tritium data
in terms of the CPT mass and mixing asymmetries
\begin{align}
A_{\Delta{m}^{2}}^{\text{CPT}}
=
\null & \null
\Delta{m}^{2}_{\nu} - \Delta{m}^{2}_{\bar\nu}
\,,
\label{004}
\\
A_{\sin^22\vartheta}^{\text{CPT}}
=
\null & \null
\sin^22\vartheta_{\nu} - \sin^22\vartheta_{\bar\nu}
\,,
\label{005}
\end{align}
we obtained
the best-fit values \cite{1008.4750}
\begin{equation}
(A_{\sin^22\vartheta}^{\text{CPT}})_{\text{bf}}
=
0.42
\,,
\quad
(A_{\Delta{m}^{2}}^{\text{CPT}})_{\text{bf}}
=
0.37 \, \text{eV}^2
\,.
\label{007}
\end{equation}
The allowed regions in the
$A_{\sin^22\vartheta}^{\text{CPT}}$--$A_{\Delta{m}^{2}}^{\text{CPT}}$ plane
are shown in Fig.~\ref{006b}.
We used a logarithmic scale for $A_{\sin^22\vartheta}^{\text{CPT}}$,
considering only the interval
$10^{-3} \leq A_{\sin^22\vartheta}^{\text{CPT}} \leq 1$
which contains all the allowed regions.
For $A_{\Delta{m}^{2}}^{\text{CPT}}$
we used an antisymmetric logarithmic scale,
which allows us to show both positive and negative values of
$A_{\Delta{m}^{2}}^{\text{CPT}}$,
enlarging the region of small values of $A_{\Delta{m}^{2}}^{\text{CPT}}$
between 0.1 and 1 eV$^2$.

The best-fit value
$(A_{\Delta{m}^{2}}^{\text{CPT}})_{\text{bf}}$
of the mass asymmetry
is small,
but Fig.~\ref{006b} shows that
in practice any value of the mass asymmetry is allowed,
with a slight preference for positive values.
On the other hand,
we obtain a very interesting result for the mixing asymmetry:
the best-fit value
$(A_{\sin^22\vartheta}^{\text{CPT}})_{\text{bf}}$
is large and positive and
Fig.~\ref{006b} shows that zero or negative values are disfavored.

\begin{figure}[t!]
\includegraphics*[bb=24 147 564 702, width=\linewidth]{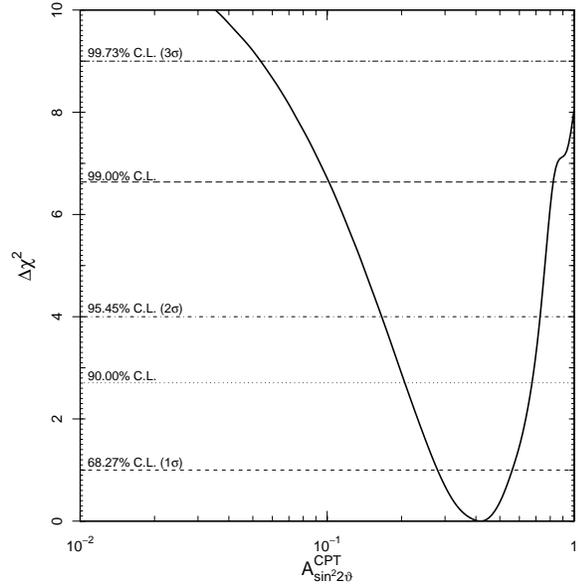}
\caption{ \label{008}
Marginal $\Delta\chi^{2}$
for
$A_{\sin^22\vartheta}^{\text{CPT}}$
\cite{1008.4750}.
}
\end{figure}

From Fig.~\ref{006b} one can see that
the smallest value of
$A_{\sin^22\vartheta}^{\text{CPT}}$
included in the $3\sigma$ allowed region
is about
0.005
at
$A_{\Delta{m}^{2}}^{\text{CPT}} \simeq -0.15 \, \text{eV}^2$.
However,
since in practice
$A_{\Delta{m}^{2}}^{\text{CPT}}$
is not bounded,
the statistically reliable limits on $A_{\sin^22\vartheta}^{\text{CPT}}$
are given by the marginal $\Delta\chi^{2}=\chi^{2}-\chi^{2}_{\text{min}}$
function for
$A_{\sin^22\vartheta}^{\text{CPT}}$
depicted in Fig.~\ref{008}.
One can see that
$A_{\sin^22\vartheta}^{\text{CPT}}>0.055$
at $3\sigma$.

The marginal $\Delta\chi^{2}$ of a null asymmetry
($A_{\sin^22\vartheta}^{\text{CPT}}=0$)
is
$12.0$,
with an associated p-value of
$0.05\%$.
Hence,
there is an indication of a positive asymmetry
$A_{\sin^22\vartheta}^{\text{CPT}}$
at a level of about $3.5\sigma$.

The indication in favor of a CPT asymmetry that we have found is robust,
because it is obtained by confronting the observations on the disappearance of electron
neutrino and antineutrino,
which should be equal if the CPT symmetry is not violated.
We considered the simplest case of a difference of
the effective squared-masses and mixings of
neutrinos and antineutrinos.
The analysis of the data in the framework of other, more complicated,
models would lead to a similar indication of a CPT asymmetry
in the space of the parameters of the specific model under consideration.

The short-baseline disappearance of electron neutrinos
can be tested in the future not only with new
Gallium radioactive source experiments,
but also with accelerator experiments with a well-known flux of electron neutrinos,
as discussed in Ref.~\cite{1005.4599}.
For the investigation of the CPT asymmetry,
the ideal experiments are those which can measure the disappearance of both
electron neutrinos and antineutrinos,
with sources which emit well-known neutrino and antineutrino fluxes
and detection processes with well-known cross sections.
Experiments of this type are near-detector
beta-beam \cite{0907.3145}
and
neutrino factory \cite{0907.5487}
experiments,
which are under study but may require a long time to be realized.
In a shorter time it may be possible to perform dedicated experiments with
intense artificial radioactive sources of electron neutrinos and antineutrinos
placed near a neutrino elastic scattering detector with a low energy threshold,
as Borexino \cite{hep-ex/9901012,Bellotti-private-10}.

\bibliography{bibtex/nu}

\end{document}